\documentclass[onecolumn,showpacs,preprintnumbers,amsmath,amssymb]{revtex4}

\usepackage{graphicx}% Include figure files
\usepackage{dcolumn}% Align table columns on decimal point
\usepackage{bm}% bold math
\usepackage{epsfig}

\begin{document}
\preprint{IST/CFP 5.2007-M J Pinheiro}

\title[]{Anomalous Diffusion at Edge and Core of a Magnetized Cold Plasma}
\author{Mario J. Pinheiro}
\address{Department of Physics and Center for Plasma Physics,\&
Instituto Superior Tecnico, Av. Rovisco Pais, \& 1049-001 Lisboa,
Portugal} \email{mpinheiro@ist.utl.pt}

\thanks{We acknowledge partial financial support from Funda\c{c}\~{a}o para a Ci\^{e}ncia e
Tecnologia and the Rectorate of the Technical University of Lisbon.}

%\subjclass{05.70.-a,52.25.Fi}

\pacs{05.70.-a,52.25.Fi,52.55.-s,55.55.Dy,52.40.Hf}

\keywords{Thermodynamics,Transport parameters,Magnetic confinement
and equilibrium,General theory and basic studies of plasma
lifetime,Plasma-material interactions}

\date{\today}
%\dedicatory{abc}%
%\commby{abc}
% ----------------------------------------------------------------
\begin{abstract}
Progress in the theory of anomalous diffusion in weakly turbulent
cold magnetized plasmas is explained. Several proposed models
advanced in the literature are discussed. Emphasis is put on a new
proposed mechanism for anomalous diffusion transport mechanism based
on the coupled action of conductive walls (excluding electrodes)
bounding the plasma drain current (edge diffusion) together with the
magnetic field flux "cutting" the area traced by the charged
particles in their orbital motion. The same reasoning is shown to
apply to the plasma core anomalous diffusion. The proposed mechanism
is expected to be valid in regimes when plasma diffusion scales as
Bohm diffusion and at high $B/N$, when collisions are of secondary
importance.
\end{abstract}

\maketitle
% ----------------------------------------------------------------

\section{Introduction}

Any phenomena occurring with interfacial systems has a fundamental
importance in science and technology~\cite{Melehy_1} (e.g.
electrostatic charging of insulators, surface tension, forward
conduction in p-n junctions). Specifically, the problem of the
plasma-wall interactions is of major importance in plasma physics.

Historically, the anomalously high diffusion of ions across magnetic
field lines in Calutron ion sources (electromagnetic separator used
by E. Lawrence for uranium isotopes) gave the firsts indications of
the onset of a new mechanism~\cite{Bohm}. It has been noticed that
the plasma moves across the magnetic confining field at a much
higher average velocity than it is predicted by classical
considerations. The classical diffusion coefficient is given by
$D_{\perp}=\eta p/B^2$, while the anomalous diffusion coefficient by
$D_{\perp}=\alpha kT/B$. Manifestly it is needed a better
understanding of the physical laws governing matter in the far
nonequilibrium state, and this is a challenging issue for the
advancement of this frontier of physics.

General proposed explanations were advanced. The first one was
Simon's "short-circuit" problem, suggesting that the observed losses
could be explained by the highly anisotropic medium induced by the
magnetic field lines, favoring electron current to the conducting
walls~\cite{Simon}. Experiments done by Geissler~\cite{Geissler1} in
the 1960's have shown that diffusion in a plasma across a magnetic
field was nearly classical (standard) diffusion when insulating
walls impose plasma ambipolarity, but in the presence of conducting
walls charged particles diffused at a much higher rate.

This problem of plasma-wall interaction becomes more complex when a
complete description is aimed of a magnetized nonisothermal plasma
transport in a conducting vessel. Beilinson {\it et
al.}~\cite{Beilinson} have shown the possibility to control the
discharge parameters by applying potential difference to sectioned
vessel conducting walls.

In the area of fusion reactors, there is strong indication that for
plasmas large but finite Bohm-like diffusion coefficient appears
above a certain range of $B$~\cite{Montgomery2}. Experiments give
evidence of transport of particles and energy to the
walls~\cite{Luce}. At the end of the 1960s, experimental results
obtained in weakly ionized plasma~\cite{Geissler1} and in a hot
electron plasma~\cite{Ferrari} (this one proposing a possible
mechanism of flute instability) indicated a strong influence that
conducting walls have on plasma losses across magnetic field lines.
Geissler~\cite{Geissler1} suggested that the most probable
explanation was the existence of diffusion-driven current flow
through the plasma to the walls. Concerning fusion reactions,
Taylor~\cite{Taylor} provided a new interpretation of tokamak
fluctuations as due to an inward particle flux resulting from the
onset of filamentary currents.

Progress in the understanding of the generation of confinement
states in a plasma is fundamental~\cite{Itoh} to pursue the dream of
a fusion reactor~\cite{Bickerton,Shafranov}. Anomalous diffusion is
a cornerstone in this quest, as recent research with tokamaks
suggest that the containment time is $\tau \approx 10^8 R^2/2D_B$,
with $R$ denoting the minor radius of a tokamak plasma and $D_B$ is
the Bohm diffusion coefficient ~\cite{Rostoker}. Controlled nuclear
fusion experiments have shown that transport of energy and particles
across magnetic field lines is anomalously large (i.e, not predicted
by classical collision theory).

The conjecture made by Bohm is that the diffusion coefficient is
$D_B=\alpha kT/eB$, where $T$ is the plasma temperature and $\alpha$
is a numerical coefficient, empirically taken to be $1/16$
~\cite{Bohm}. Initially, the origin of the anomalous diffusion was
assumed to be due to the turbulence of small-scale instabilities
(see, for example, Refs.~\cite{Montgomery2,Taylor,Montgomery1}).
However, it is now clear that there is a number of different
mechanisms that can lead to anomalous diffusion such as, coherent
structures, avalanches type processes and streamers, which have a
different character than a purely diffusive transport process.
Recent experimental results such as scaling of the confinement time
in L-mode plasmas and perturbative experiments undermine the
previous paradigm built on the standard transport
processes~\cite{Callen,Berk} showing conclusively that there are
many regimes where plasma diffusion does not scale as $B^{-1}$.

This paper put emphasis on a mechanism of wall current drain set up
together with the magnetic field "cutting" lines across the area
traced by the charged particles trajectories. The proposed mechanism
of anomalous diffusion is expected to be valid in purely diffusive
regimes when plasma diffusion scales as Bohm diffusion, both in the
edge and core of a cold magnetized plasma. At his stage it was
considered of secondary importance the role of collisions in
randomizing the particle's distribution function. From collisional
low temperature plasmas to a burning fusion plasma subject the
plasma confinement vessel to strong wall load, both in stellarator
or tokamak operating modes, this explanation could be of
considerable interest, particularly when diffusive transport process
are dominant.

We would like to stress that this work is not free from omission of
important contributions.

\section{Simon's "short-circuit" theory}

The first attempt to explain why the plasma diffuse at a much higher
average velocity than it is predicted by classical theory has been
advanced by A. Simon~\cite{Simon}. The magnetic field lines
structure a highly anisotropic medium. Any fluctuation of the space
charge builds up an electric field, which has a strong effect on the
currents parallel to the magnetic field lines. In fact, the
classical equation for conductivity across a magnetic field is given
by
\begin{equation}\label{}
\sigma_{\perp} = \frac{\sigma_0}{1+\frac{\Omega_c^2}{\nu_e^2}},
\end{equation}
where $\Omega_c=eB/m$ is the electron cyclotron frequency, $\nu_e$
is the electron collision frequency, and $\sigma_0=e^2 n_e/m\nu_e$
is the conductivity in the absence of a magnetic field. By the
contrary, due to $\Omega_c/\nu_e \gg 1$, this electric field is too
small to have any importance on the crossed-field conductivity. From
this results that there is a strong current to the wall without a
concomitant current to different regions of the plasma, making of
this situation a kind of circuital "short-circuit" problem. Although
Simon attempted to explain the anomalously high rate of diffusion in
Calutron ion sources in the frame of the classical diffusion theory
calculating the coefficient $D_{\perp}$ as being approximately equal
to the transverse diffusion coefficient of the ions. His proposal is
not suitable, however, because the experimental determination of
$D_{\perp}$ by means of a decaying plasma have shown that according
to the magnetic field strengthen the transverse diffusion
coefficient can be much higher than the classical one or smaller
than the transverse diffusion coefficient of the
ions~\cite{Geissler1}.

\section{Plasma turbulence and transport}

Purely diffusive transport models cannot give convincing
explanations for a variety of experiments in magnetically confined
plasmas in fusion engineering devices, particularly the scaling of
the confinement time in L-mode plasmas. The assumed underlying
instabilities are driven by either the pressure gradient or the ion
temperature gradient. It is well established fact that transport in
high temperature confined plasmas is driven by turbulence and plasma
profiles, and are subject to transition from L-mode to H-mode
(characterized by a a very steep gradient near the plasma
surface)~\cite{Itoh_03}. The non-linearity in the gradient-flux
relation is the source of turbulence and turbulence-driven
transport. The fluxes contain all the dynamic information on the
transport process. Accordingly, changes in the gradient trigger
local instabilities in the plasma. This local instability induces an
increase in the nearby gradients, thus causing a propagation of the
instability all across the plasma. In particular, an excessive
pressure on the core propagates to the edge in a kind of avalanche.

In weakly turbulent cold magnetized plasmas, besides the Calutron
ion sources and the magnetron, the study of particle transport in
crossed electric and magnetic fields results from applications to
electromagnetic space propulsion (Hall thrusters). Those plasma
accelerators work with a radial magnetic field that prevents
electron flow toward the anode and forcing the electrons closed-loop
drift around the axis of the annular geometry. Neutrals coming from
the anode are ionized in this rotating electrons cloud, while ions
are accelerated by an axial electric field that freely accelerate
them out from the device. This effect develops in the so called
extended acceleration zone (or electric-magnetic region plasma). In
this acceleration zone the electron gyro-radius and the Debye
shielding length are small relative to the apparatus dimension,
while the ion gyro radius is larger than the apparatus typical
length. From these spatial scales results that the electron motions
are $[\mathbf{E} \times \mathbf{B}]$ drifts, but ions are
accelerated by the electric field that develops in the plasma. The
first observations of a large amount of electron transport toward
the anode have been noticed in the 60's (e.g., Ref.~\cite{Janes_66})
and they have been related to electric field fluctuations since they
were correlated with the density variations in order to produce
anomalous transport. Other possible mechanism that could explain the
high electrons transverse conductivity were advanced: collisions
with the wall~\cite{Morozov_72,Morozov_87}. Electrons moving freely
along the lines of forces of the magnetic field collide with the
wall more frequently than with ions and neutrals, being reflected at
the wall and enhancing emission of low-energy secondary electrons
from the wall. As referred, the other strong candidate, which could
possibly be the source of a higher axial electron current than
predicted by the standard classical kinetic theory, is the turbulent
plasma fluctuations. But it seems that there is no clear consensus
on this issue~\cite{Boeuf_98,Smirnov_04,Hofer_06}.

The magnetron is a sputtering tool, used for reactive deposition and
etching. The magnetron effect is applicable to different geometries
and only need a closed-loop $[\mathbf{E} \times \mathbf{B}]$ drift
to work. Rossnagel {\it et al.}~\cite{Rossnagel_86} have shown that
the Hall-to-discharge current ratio measured in those configurations
could be explained if the high collision frequencies for electrons
were associated to Bohm diffusion. In particular,
Kaufman~\cite{Kaufman_85} argues that anomalous diffusion in
closed-loop $[\mathbf{E} \times \mathbf{B}]$ thrusters could shift
from core diffusion to edge diffusion (or wall effects) with
increasing magnetic fields.

\section{Circuital model of anomalous diffusion}

\begin{figure}
  \includegraphics[width=3.0 in, height=3.5 in]{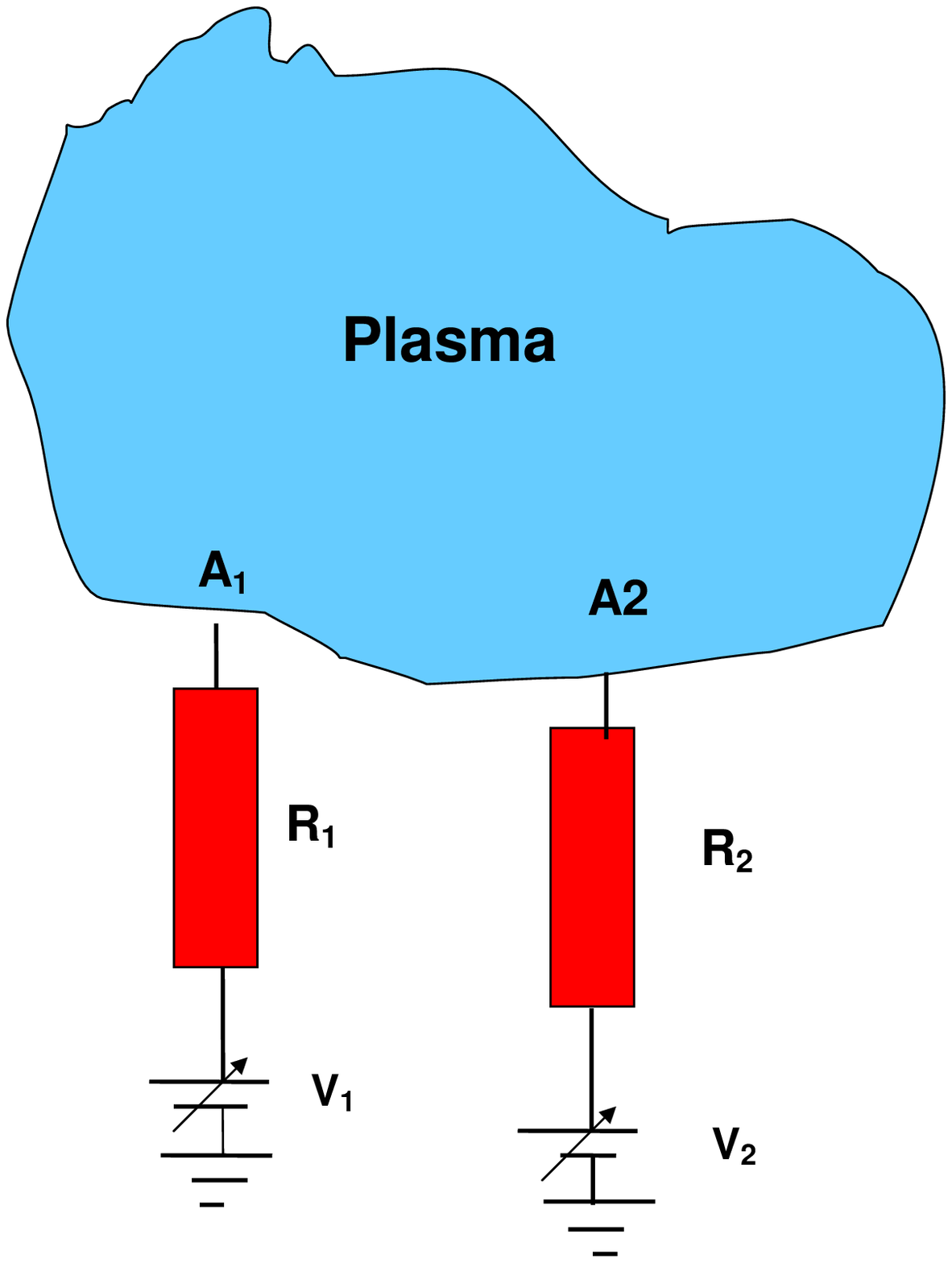}\\
  \caption{Schematic diagram of the plasma boundary connected to the common circuit through
  conducting walls.}\label{Fig1}
\end{figure}

In a seminal paper~\cite{Robertson} a conjecture was proposed based
on the principle of minimum entropy-production rate, stating that a
plasma will be more stable whenever the internal product of the
current density $\textbf{j}$ by an elementary conducting area $d
\mathcal{A}$ at every point of the boundary - excluding the surface
collecting the driving current - is null, $(\mathbf{j} \cdot
\mathcal{A})=0$ at any point of the boundary (and excluding the
surfaces collecting the discharge current), independently of the
resistance $R_i$. The general idea proposed by
Robertson~\cite{Robertson} assumes that the plasma boundary is
composed of small elements of area $\mathcal{A}_i$, each one
isolated from the others, but each one connected to the exterior
common circuit through its own resistor $R_i$ and voltage $V_i$. The
entropy production rate in the external circuits is given by:
\begin{equation}\label{Eq1}
\frac{d S}{d t} = \sum_i \frac{1}{T} (\mathbf{j}_i \cdot
\mathcal{A}_i)^2 R_i,
\end{equation}
where $T$ is the temperature of the resistors, supposed to be in
thermal equilibrium with all the others. It is important to remark
that the summation is over the different conducting areas eventually
confining the plasma, {\it excluding} the electrodes areas.
Fig.\ref{Fig1} illustrates this concept.

We consider a simple axysymmetric magnetic configuration with
magnetic field lines parallel to z-axis with a plasma confined
between two electrodes (see Fig.1). In general terms, a particle's
motion in the plasma results in a massive flux. As long as the flux
is installed, the flux will depend naturally on a force $\mathbf{F}$
- in this case the pressure gradient-driven process of diffusion to
the wall - responsible of the wall driven current $\mathbf{j}$.
According to the fundamental thermodynamic relation, the plasma
internal energy variation $dU$ is related to the amount of entropy
supplied or rejected and the work done by the driven force, through
the equation:
\begin{equation}\label{Eq5}
\frac{dU}{dt} = (\mathbf{j} \cdot \mathcal{A})^2 R + \left(
\mathbf{F} \cdot \frac{d \mathbf{r}}{dt} \right).
\end{equation}

The last term we identify with the macroscopic diffusion velocity
$\mathbf{v}_d$ depicting the process of plasma expansion to the
wall. To simplify somehow the calculations we assume a single plasma
fluid under the action of a pressure gradient
($\mathbf{F}=\mathcal{A} L dp/dy \overrightarrow{j}$, where
$\overrightarrow{j}$ is a unit vector directed along the Oy axis).

In the presence of steady and uniform magnetic field lines (this
simplifies the equations, but do not limit the applicability of the
model), the particles stream freely along them. From
magnetohydrodynamics we have a kind of generalized Ohm's law (see,
for example, Ref.~\cite{Kadomtsev1}):
\begin{equation}\label{Eq6}
\nabla p = - e n \mathbf{E} - e n [\mathbf{v} \times \mathbf{B}] +
[\mathbf{j} \times \mathbf{B}] - \frac{e n \mathbf{j}}{\sigma},
\end{equation}
where $\sigma = e^2 n \tau_e/m_e$ is the electric conductivity, with
$\tau_e$ denoting the average collision time between electrons and
ions. The force balance equation is given by:
\begin{equation}\label{Eq7}
\nabla p = [\mathbf{j} \times \mathbf{B}],
\end{equation}
valid whenever the Larmor radius is smaller than the Debye radius.
This assumption simplifies further the extension of our model to
high enough magnetic fields. Therefore, after inserting
Eq.~\ref{Eq7} into Eq.~\ref{Eq6} the y component of velocity is
obtained:
\begin{equation}\label{Eq8}
v_y = - \frac{E_x}{B} - \frac{1}{\sigma B^2} \frac{d p}{d y}.
\end{equation}
From Eq.~\ref{Eq8} we have the classical diffusion coefficient
scaling with $1/B^2$ and thus implying a random walk of step length
$r_L$ (Larmor radius). To get the anomalous diffusion coefficient
and as well understand better its related physics, we must consider
the process of diffusion to the wall - in the presence of an entropy
source - with the combined action of the wall current drain, as
already introduced in Eq.~\ref{Eq5}.

Therefore, using the guiding center plasma model the particle motion
is made with velocity given by:
\begin{equation}\label{Eq9}
\mathbf{j}=en\mathbf{v}_d=-\frac{[\nabla p \times
\mathbf{B}]}{B^2}.
\end{equation}
This equation forms the base of a simplified theory of magnetic
confinement. In fact, the validity of Eq.~\ref{Eq9} is restrained to
the high magnetic field limit, when the Larmor radius is shorter
than the Debye radius.

Considering the motion along only one direction perpendicular to the
wall (y-axis), it is clear that
\begin{equation}\label{Eq10}
(\mathbf{j} \cdot \mathcal{A})^2 = \frac{\mathcal{A}^2}{B^2}
\left( \frac{dp}{dy} \right)^2.
\end{equation}
If we consider a quasi-steady state plasma operation, the plasma
total energy should be sustained. Hence, $dU/dt=0$, and the power
associated with the driven pressure-gradient is just maintaining the
dissipative process of plasma losses on the wall. Eq.~\ref{Eq5}
governs the evolution of the diffusion velocity. Hence, we have
\begin{equation}\label{Eq12}
n v_d = - \frac{n R \mathcal{A}}{L} \frac{kT}{B^2} \frac{dn}{dy} =
-D_{T} \frac{dn}{dy},
\end{equation}
with $D_T$ denoting the transverse (across the magnetic field)
diffusion coefficient given by:
\begin{equation}\label{Eq13}
D_T = \frac{n R \mathcal{A}}{L} \frac{kT}{B^2}.
\end{equation}
This new result coincides with the classical diffusion
coefficient~\cite{Roth} whenever $nR\mathcal{A}/L \equiv m
\nu_{ei}/e^2$, containing a dependence on collision frequency and
particle number density. Other theoretical approaches to this
problem were advanced by Bohm~\cite{Bohm}, who proposed an
empirically-driven diffusion coefficient associating plasma
oscillations as the source of the enhanced diffusion, while
Tonks~\cite{Tonks} have shown that the current density that is
present in a magnetically immobilized plasma is only generated by
the particle density gradient, not being associated with any drift
of matter. Simon electron "short-circuit"~\cite{Simon} scheme
attempt to explain the different rates of diffusion, electrons and
ions do experiment across the magnetic field. While the ion flux
dominates the radial diffusion, the electron flux dominates axial
losses, due to an unbalance of currents flowing to the wall.

In the absence of collisions, the guiding centers of charged
particles behave as permanently attached to the same lines of
force. On the contrary, as a result of collisions with others
charged particles the guiding centers shift from one line of force
to another resulting in a diffusion of plasma across the field
lines. In our model, each orbit constitutes an elementary current
$I$ eventually crossing the wall.

\begin{figure}
  \includegraphics[width=3.5 in, height=4.5 in]{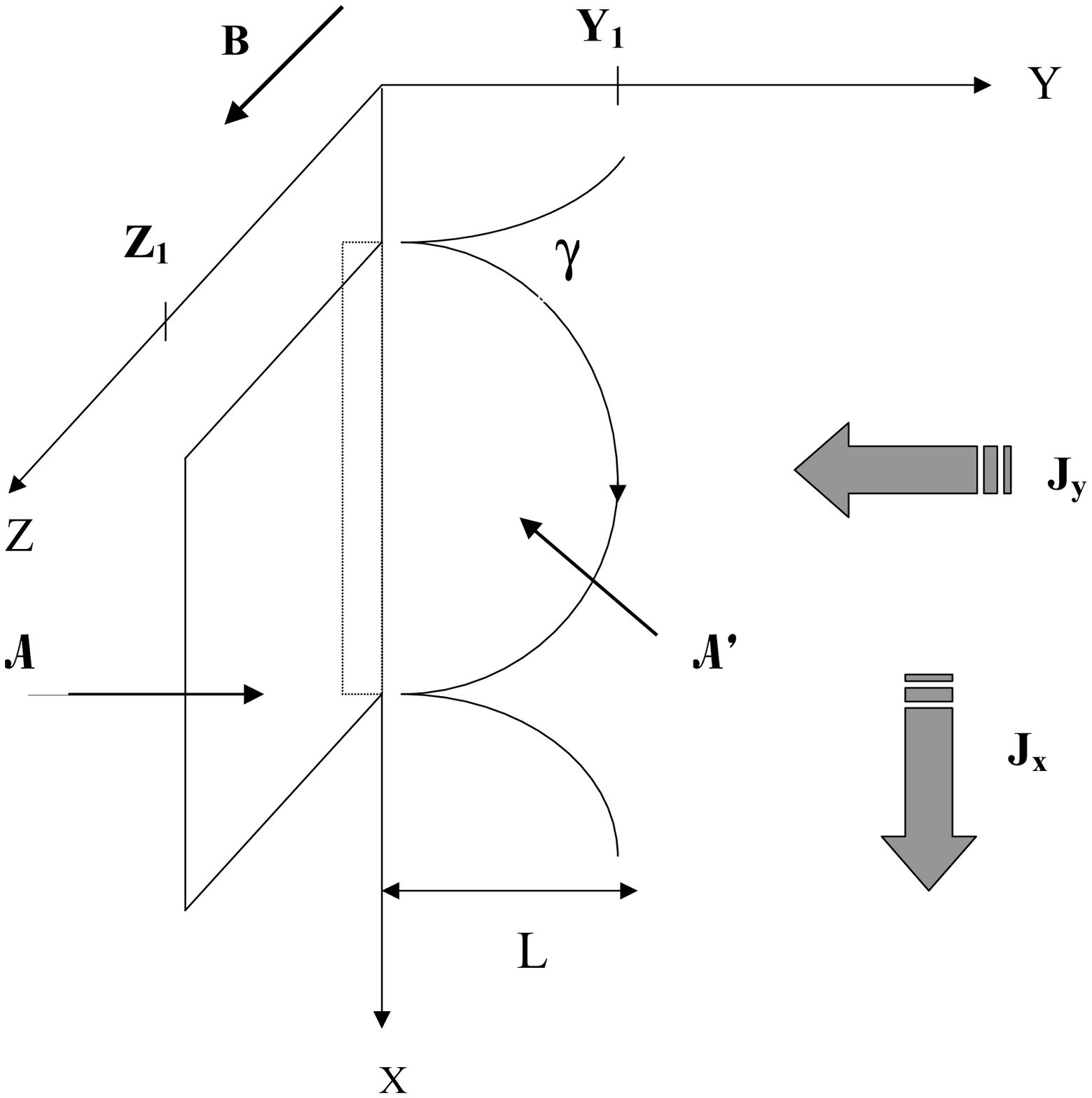}\\
  \caption{Schematic of the geometry for the plasma-wall current drain model. The uniform magnetic field
  points downward along Oz. Particles describe orbits in the plane xOy intersecting the wall
  (plan xOz). Orbits are represented by a semi-circular line for convenience.
  $L$ is the maximum distance the trajectory attains
  from the wall.}\label{fig1}
\end{figure}

However, the particle diffusion coefficient as shown in
Eq.~\ref{Eq13} gives evidence of an interplay with the resistance
that the elementary circuit offer when in contact with the walls in
the presence of the frozen-in effect. In fact, for sufficiently
strong magnetic fields apparently a hydrodynamic behavior of the
plasma is installed ~\cite{Montgomery2,Corkum}, with the appearance
of "convective cells" and the $1/B$ behavior dominates, giving birth
to the anomalous diffusion mechanism. The onset of freezing magnetic
lines is valid whenever the Lundquist number $\mathrm{S} \gg 1$
(convection of the magnetic field dominated medium). In this case
the magnetic field lines are frozen-in in the medium (consequence of
a vortex type of character of the magnetic field $\mathbf{B}$) and
the flux of them across a given surface is constant:
\begin{equation}\label{Eq14}
\Phi =B \mathcal{A}' = B L^2 \alpha.
\end{equation}
Remark that $\mathcal{A}'$ is now the surface delimited by the
elementary circuit $\gamma$ (see Fig.~\ref{fig1}) and $\alpha
\lesssim 1$ is just a geometrical factor (e.g. $\alpha=\pi/4$ at the
limit of a circular orbit). This situation is fundamental to the
onset of anomalous diffusion. Free electrons orbits are helical, but
as Fig.~\ref{fig1} shows, their projections at rigth angles to the
field are circular. Each particle orbit constitute an elementary
circuit with $B$-field cutting its surface being associated with it
an elementary flux $\Phi$. At the same time we can envisage each
orbit as constituting by itself an elementary circuit, some of them
intersecting the wall and thus the circuit is closed inside the
wall. Therefore a resistance $R$ drags the charged flow at the
conducting wall. It is therefore plausible to associate to this
elementary circuit a potential drop $V$ and all the process being
equivalent to a current $I$ flowing through the elementary circuit.

Assuming the plasma is a typical weakly coupled, hot diffuse
plasma with a plasma parameter (number of particles in Debye
sphere) $\Lambda = n\lambda_{De}^3 \approx 1$, it is more likely
to expect nearly equal average kinetic and potential energy.
However, the typical plasma parameter encountered in glow
discharges or in nuclear fusion is $\Lambda \gg 1$. This means
that the average kinetic energy is larger than the average
potential energy. To contemplate all range of $\Lambda$ we can
relate them through the relationship
\begin{equation}\label{Eq15}
\rho V = (\mathbf{J} \cdot \mathbf{A}) \delta.
\end{equation}
Here, $\rho$ is the charge density, $\mathbf{A}$ is the vector
potential, $\mathbf{J}$ is the current density and $\delta \leq 1$
is just a parameter representing the ratio of potential to kinetic
energy. Of course, when $\Lambda \geq 1$, then $\delta \leq 1$.
This basic assumption is consistent with the hydrodynamic
approximation taken in the development of equations. The
limitations of the model are related with the unknowns $\Lambda$
and $\delta$ that can be uncovered only through a self-consistent
model of the plasma. However, our analysis of anomalous diffusion
remains general and added new insight to the phenomena.

Now suppose that the diffusion current is along y-axis
$\mathbf{J}=-J_y \mathbf{u}_y$ (see Fig.1). Consequently,
$\mathbf{A}=-A_y \mathbf{u}_y$, and then the potential drop will
depend on x-coordinate:
\begin{equation}\label{Eq16}
\rho [V(x_1) - V(x_0)] = J_y [A_y(x_1) - A_y(x_0)] \delta.
\end{equation}
Multiplying both members by the area $\mathcal{A}'=x_1 z_1$ and
length $L=y_1$, we have
\begin{equation}\label{Eq17}
Q \Delta V = I y_1 [A_y(x_1) - A_y(x_0)] \delta =I \Phi \delta.
\end{equation}
$\Phi=\oint_{\gamma} (\mathbf{A} \cdot d\mathbf{x})$ is the flux
of the magnetic field through the closed surface bounded by the
line element $d \mathbf{x}$ (elementary circuit $\gamma$, see also
Fig.\ref{fig1}). By other side, naturally, the total charge
present on the volume $\mathcal{V}=x_1 y_1 z_1$ is such as $Q=ie$,
with $i$ an integer. This integer must be related to ions charge
number. From Eq.~\ref{Eq17} we obtain
\begin{equation}\label{Eq19}
R = \frac{\Delta V}{I} = \delta \frac{\Phi }{Q} = \alpha \delta
\frac{B L^2 }{i e}.
\end{equation}
But, the particle density is given by $n=N/L\mathcal{A}$, with $N$
being now the total number of charged particles present in volume
$\mathcal{V}=\mathcal{A}L$. Since $i=N$, we retrieve finally the
so-called Bohm-diffusion coefficient
\begin{equation}\label{Eq18}
D_B = \alpha \delta \frac{kT}{eB}.
\end{equation}

\begin{figure}
  \includegraphics[width=3.5 in, height=3.5 in]{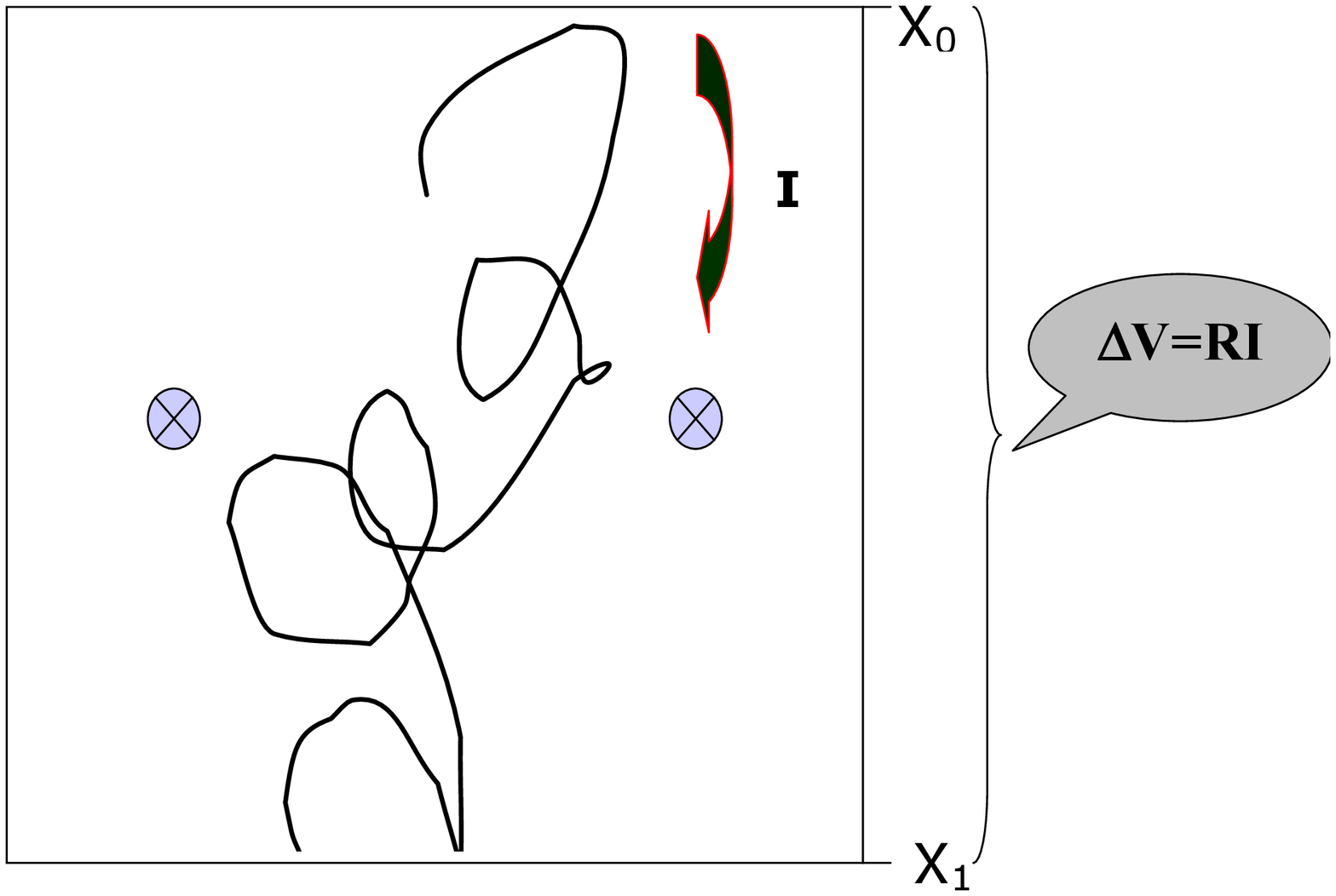}\\
  \caption{Volume control and particle's trajectory submitted to magnetic field.
  Magnetic field lines point downward.}\label{fig3}
\end{figure}

So far, our arguments were applied to edge anomalous diffusion. But
they can be generalized to the core anomalous diffusion processes,
provided that diffusive transport processes are dominant. For this
purpose consider instead of a conducting surface a virtual surface
delimiting a given volume, as shown in Fig.~\ref{fig3}.

\begin{figure}
  \includegraphics[width=3.5 in, height=3.0 in]{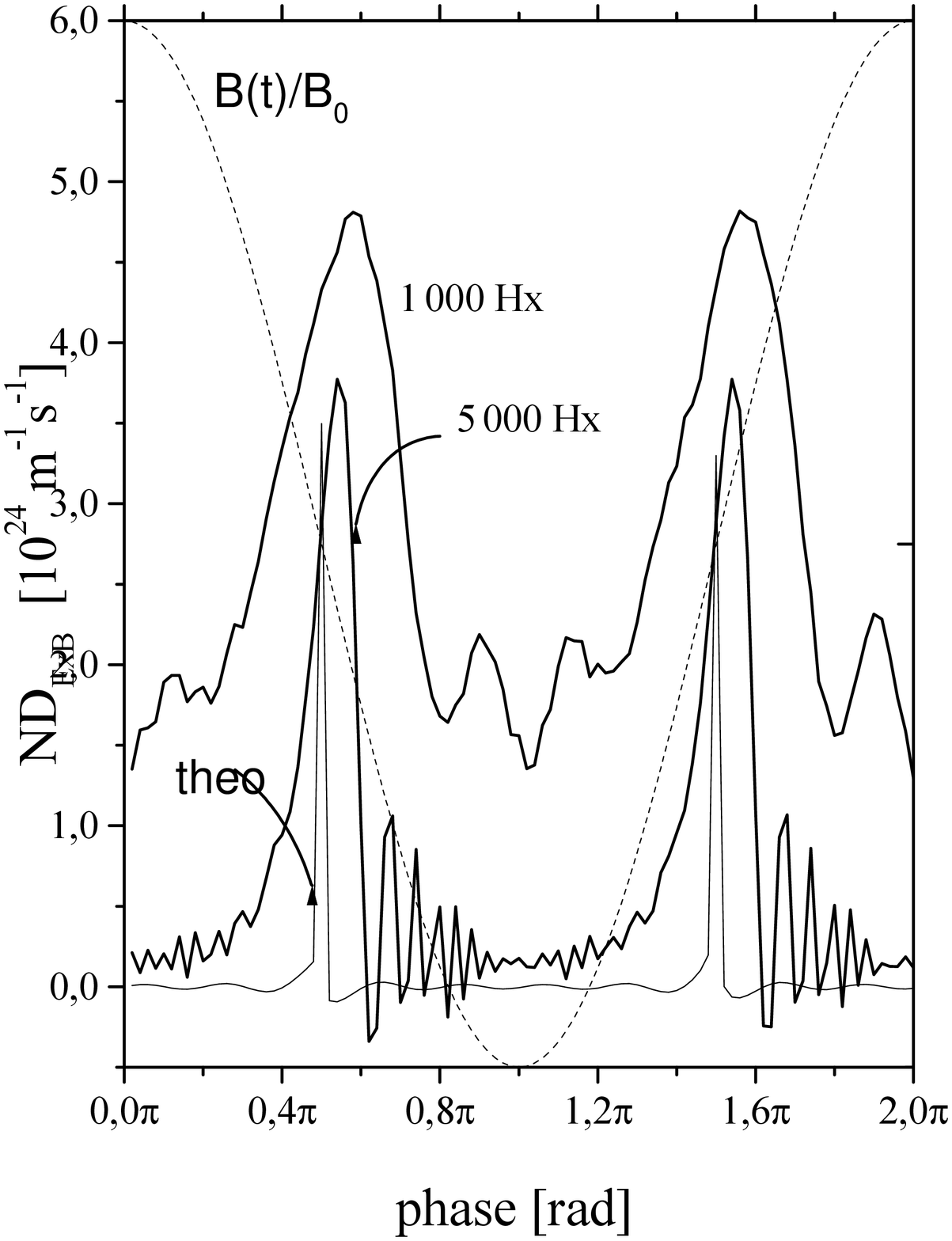}\\
  \caption{Comparison between numerical results for 5000 Hx obtained by Monte Carlo simulations Refs.[32,33]
  of electron transport in crossed electric and magnetic rf fields in argon and Eq.~\ref{formulafinal}
  -theo. Dashed line: external time-dependent magnetic field. Parameters used: $\delta=\frac{1}{40}$; $\chi$;
  applied frequency, f=100 MHz; ciclotron frequency,
  $\Omega_c \approx 10^{10}$ Hz; $\frac{kT}{e}=5.4$ eV; p=1 Torr, $T_g=300$ K, $\chi_=0$.}\label{fig4}
\end{figure}

Our coefficient is time-dependent and can be written under the form:
\begin{equation}\label{}
D_{\perp}=\delta \frac{kT(t)}{eB(t)^2}\frac{\Phi(t)}{L^2}.
\end{equation}
The nonrelativistic solutions of dynamical equation of a charged
particle in time-dependent but homogeneous electric and magnetic
field give the following approximative expressions for the width
trajectories along the xOy plane (see, e.g.
Ref.\cite{chandrasekhar_60}) by:
\begin{equation}\label{}
\begin{array}{cc}
  \Delta x = \frac{v_{\perp}}{\Omega_c}\cos(\Omega_c t + \chi) & \Delta
  y=\mp \frac{v_{\perp}}{\Omega_c} \sin(\Omega_c t + \chi),
\end{array}
\end{equation}
where $\chi$ is the initial phase, $v_{\perp}$ denotes the component
of the velocity perpendicular to the magnetic field and the $\mp$
sign applies to electrons (-) or positive ions (+). From them we can
retrieve the flux "cutting" area:
\begin{equation}\label{}
\mathcal{A} \approx \Delta x.\Delta y =-\frac{1}{2} \left(
\frac{\nu_{\perp}}{\Omega_c} \right)^2 \sin(2 \Omega_c t + 2 \chi).
\end{equation}
Then the anomalous diffusion coefficient is just given by:
\begin{equation}\label{formulafinal}
D_{\perp} \approx \mp \delta
\frac{kT(t)}{eB(t_0)}\frac{1}{2}\frac{\sin(2 \Omega_c t +
2\chi)}{\mid \cos(\omega t) \mid}.
\end{equation}
As we can see in Fig.\ref{fig4} this last expression describes
fairly well the diffusion process for high enough $B/N$ values (the
magnetic field to gas number density ratio) and explains the main
processes building-up such effects as: i) the negative diffusion,
which results from the contraction of the flux "cutting" area; ii)
the ciclotronic modulation imprint on the transverse diffusion
coefficient; iii) and the anomalous diffusion, due to the fast flux
rate of the magnetic field through the area $\mathcal{A}$). All this
signs can be seen on Fig.~\ref{fig4} were it is shown a comparison
of numerical results (5000 Hx, 1 Hx=$10^{-27}$ T.m$^3$) obtained
with Monte Carlo simulations of electron transport in crossed
magnetic and electric fields by Petrovi\'{c} {\it et
al.}~\cite{Zoran,Zoran1} with the theoretical prediction given by
our Eq.~\ref{formulafinal}. As long as only a self-consistent model
could give us an exact value of the ratio of potential to kinetic
energy $\delta$, we assume here $\delta=1/40$. The full agreement
with the numerical calculations is not obtained due to neglecting
effects related to the electric field variation in time and of the
assumed collisionless approximation. This explains the big
discrepancy shown in Fig.~\ref{fig4} when compared with the
diffusion coefficient at 1000 Hx when collisions begin to be far
more important to randomize individual trajectories and our approach
is no more valid (at low enough $B/N$ values).

\section{Discussion and Summary}

However, Eq.~\ref{Eq18} suffers from the indetermination of the
geometrical factor $\alpha$. This factor is related to the ions
charge number, it depends on the magnetic field magnitude and as
well as on the external operating conditions (due to increased
collisional processes, for ex.). The exact value of the product
$\alpha \delta$ can only be determined through a self-consistent
plasma model, but we should expect from the above discussion that
$\alpha \delta < 1$. For a 100-eV plasma in a 1-T field, we obtain
$D_B=1.67$ m$^2/$s (using the thermal to magnetic energy ratio with
particle's density $n=10^{14}$ cm$^{-3}$). Furthermore,
Eq.~\ref{Eq19} can be used as a boundary condition (simulating an
electrically floating surface) imposed when solving Poisson
equation.

Also it worth to emphasize that when inserting Eq.~\ref{Eq19} into
Eq.~\ref{Eq13}, and considering the usual definition of momentum
transfer cross section, then it can be obtained a new expression
for the classical diffusion coefficient as a function of the ratio
of collisional $\nu$ and cyclotron frequency $\Omega$, although
(and in contrast with the standard expression), now also dependent
on the geometrical factor $\alpha$ and energy ratio $\delta$:
\begin{equation}\label{}
D_T = (\alpha \delta) \frac{\nu}{\Omega} \frac{kT}{m}.
\end{equation}
This explains the strong dependence of the classical diffusion
coefficient on $\nu/\Omega$ showing signs of anomalous diffusion as
discussed in Ref.~\cite{Zoran} (obtained with a time resolved Monte
Carlo simulation in an infinite gas under uniform fields) and, in
addition, the strong oscillations shown up in the calculations of
the time dependence of the transverse component of the diffusion
tensor for electrons in low-temperature rf argon plasma. Those basic
features result on one side from its dependence on $R$, which is
proportional to the flux. Therefore, a flux variation can give an
equivalent effect to the previously proposed mechanism: whenever a
decrease (or increase) in the flux is onset through time dependence
of electric and magnetic fields, it occurs a strong increase (or
decrease) of the diffusion coefficient. By other side, when the
resistance increases it occurs a related decrease of charged
particles tangential velocity and its mean energy. So far, this
model gives a new insight into the results referred in~\cite{Zoran}
and also explains why the same effect is not obtained from the
solution of the non-conservative Boltzmann equation as applied to an
oxygen magnetron discharge with constant electric and magnetic
fields~\cite{White}.

A further application of Eq.~\ref{Eq1} to a cold plasma can give a
new insight into the "ambipolar-like" diffusion processes.
Considering just one conducting surface (besides the electrodes
driving the main current into the plasma) and the plasma build-up of
electrons and one ion component to simplify matters, we obtain:
\begin{equation}\label{Eq2}
\frac{d S}{dt} = \frac{e^2}{T} (-n_e \mu_e \mathbf{E} + D_e \nabla
n_e + n_i \mu_i \mathbf{E} - D_i \nabla n_i)^2 \mathcal{A}^2 R.
\end{equation}
Under the usual assumptions of quasi-neutrality and quasi-stationary
plasma (see, for example, Ref. ~\cite{Roth}), the following
conditions must be verified:
\begin{equation}\label{Eq3}
\begin{array}{cc}
  \frac{n_i}{n_e}=\epsilon=const. ; & n_e \mathbf{v_e}=n_i
  \mathbf{v}_i,
  \\
\end{array}
\end{equation}
and hence, Eq.~\ref{Eq1} becomes:
\begin{equation}\label{Eq4}
\frac{dS}{dt}=\frac{e^2}{T} [\mathbf{E}(\epsilon \mu_i - \mu_e)n_e +
\nabla n_e (D_e - D_i \epsilon)]^2 \mathcal{A}^2 R.
\end{equation}
For a stable steady-state plasma with no entropy sources the
condition $\dot{S}=0$ prevails and then an "ambipolar-like" electric
field is recovered~\cite{Roth}:
\begin{equation}\label{amb1}
\mathbf{E}=\frac{D_e - \epsilon D_i}{\mu_e - \mu_i
\epsilon}\frac{\nabla n_e}{n_e}.
\end{equation}
It means that the conducting surface must be at its floating
potential. Such conceptual formulation provides new insight into
"ambipolar-like" diffusion processes. In a thermal equilibrium
state, a plasma confined by insulating walls will have an effective
coefficient given by the above Eq.~\ref{amb1}, a situation
frequently encountered in industrial applications. This example by
itself relates ambipolar diffusion with no entropy production in the
plasma. However, allowing plasma currents to the walls, entropy
production is greatly enhanced generating altogether instabilities
and plasma losses~\cite{Robertson}. As long as confined plasmas are
in a far-nonequilibrium state (with external surroundings) it is
necessary to establish a generalized principle that rule matter, and
this circuital model for anomalous diffusion represents some
progress in the physics of plasmas as nonequilibrium systems.

To summarize, we introduced in this study a simple circuital
mechanism providing an interpretation of the anomalous diffusion in
a magnetized confined plasma in a purely diffusive transport regime.
The coupled action of the magnetic field "cutting" flux through the
areas traced by the charge carriers elementary orbits, together with
the elementary electric circuit constituted by the charged particle
trajectory itself are at the basis of the anomalous diffusion
process. Whenever conducting walls bounding the plasma drain the
current (edge diffusion) or, at the plasma core, the magnetic field
flux through the areas traced by the charged particles varies, a
Bohm-like behavior of the transverse diffusion coefficient can be
expected. Eq.~\ref{formulafinal} can be used as an analytical
formula when simulating plasma behavior at high $B/N$. In the near
future we hope to generalize this model taking into account random
collisions. The suggested mechanism could lead to a better
understanding of the mechanism of plasma-wall interaction and help
to develop a full-scale numerical modeling of present fusion devices
or collisional low-temperature plasmas.

The author gratefully acknowledge the data supplied to us by Zoran
Lj. Petrovi\'{c} and Zoran Raspopovic used in our Fig.4 .

% ----------------------------------------------------------------
%\INPUT{Xbib.bib}   % For Gather Purpose Only
%\INPUT{Doc2.bbl}  % For Gather Purpose Only
\bibliographystyle{amsplain}
\bibliography{Doc2}

\begin{thebibliography}{1}

\bibitem{Melehy_1} Mahamoud A. Melehy, AIP Conference Proceedings
{\bf 861} 524 (2006)

\bibitem{Bohm} Bohm, Burhop and Massey, {\it Characteristics of Electrical Discharges in Magnetic Fields}, edited by
 A. Guthrie and R. K. Wakcrling (MacGraw-Hill, New York,1949)

\bibitem{Simon} Albert Simon, Phys. Rev. {\bf 98} (2) 317 (1955)

\bibitem{Geissler1} Klaus H. Geissler, Phys. Rev. {\bf 171}(1)
179 (1968)

\bibitem{Beilinson} L. L. Beilinson, V. A. Rozhansky, and L. D.
Tsendin, Phys. Rev. E {\bf 50} (4) 3033 (1994)

\bibitem{Montgomery2} David Montgomery, C.-S. Liu, and George
Vahala, Phys. Fluids {\bf 15} (5), 815 (1972)

\bibitem{Luce} T. C. Luce, C. C. Petty, and J. C. M. de Haas,
Phys. Rev. Lett. {\bf 68} (1) 52 (1992)

\bibitem{Ferrari} L. A. Ferrari and A. F. Kuckes, Phys. Fluids
{\bf 12} 836 (1969)

\bibitem{Taylor} J. B. Taylor and B. McNamara, Phys. Fluids {\bf
14} (7) 1492 (1971)

\bibitem{Itoh} Kimitaka Itoh, Sanae-I. Itoh, Atsushi Fukuyama and
Masotoshi Yagi, J. Plasma Fusion Res. {\bf 79} (6) 608 (2003)

\bibitem{Bickerton} R. J. Bickerton, Phil. Trans. R. Soc. Lond. A
{\bf 375} 397 (1999)

\bibitem{Shafranov} V. D. Shafranov, Physics-Uspekhi {\bf 44} (8)
835 (2001)

\bibitem{Rostoker} Norman Rostoker, Michl W. Binderbauer, Hendrik
J. Monkhorst, Science {\bf 278} 1419 (1997)

\bibitem{Montgomery1} David Montgomery and Frederick Tappert,
Phys. Fluids {\bf 15} (4) 683 (1972)

\bibitem{Callen} J. D. Callen and M. W. Kissick, Plasma Phys.
Control. Fusion {\bf 39} B173-B188 (1997)
\bibitem{Berk} H. L. Berk, B. N. Breizman, and Huanchun Ye, Phys.
Rev. Lett. {\bf 68} (24) 3563 (1992)

\bibitem{Itoh_03} Kimitaka Itoh, Sanae-I Itoh, Atsushi Fukuyama and
Masatoshi Yagi, J. Plasma Fusion Res. {\bf 79} (6) 608 (2003)

\bibitem{Janes_66} G. S. Janes and R. S. Lowder, Phys. Fluids {\bf
9} (6) 1115 (1966)

\bibitem{Morozov_72} A. I. Morozov, Yu. V. Esinchuk, G. N. Tilinin,
A. V. Trofimov, Yu. A. Sharov, and G. Ya. Shchepkin, Sov.
Phys.-Techn. Phys. {\bf 17} (1) 38 (1972)

\bibitem{Morozov_87} A. I. Morozov, Sov. Phys. Tech. Phys. {\bf 32}
(8) 901 (1987)

\bibitem{Boeuf_98} J. P. Boeuf and L. Garrigues, J. Appl. Phys. {\bf
84} (7) 3541 (1998)

\bibitem{Smirnov_04} A. Smirnov, Y. Raitses, and N. J. Fisch, Phys.
Plasmas {\bf 11} (11) 4922 (2004)

\bibitem{Hofer_06} Richard, R. Hofer, Ira Katz, Ioannis G.
Mikellides, and Manuel Gamero-Casta\~{n}o, in {\it Proceedings of
the 42$^{nd}$ Joint Propulsion Conference, Sacramento, CA, 2006},
AIAA 2006-4658

\bibitem{Rossnagel_86} S. M. Rossnagel and H. R. Kaufman, J. Vac.
Sci. Technol. A {\bf 5} (1) 88 (1986)

\bibitem{Kaufman_85} H. R. Kaufman, AIAA J. {23} 78 (1985)






\bibitem{Robertson} Harry S. Robertson, Phys. Rev. {\bf 118} (1)
288 (1969)

\bibitem{Kadomtsev1} B. B. Kadomtsev, {\it Ph\'{e}nom\`{e}nes
collectifs dans les plasmas} (Mir Editions, Moscow, 1979)

\bibitem{Roth} J. Reece Roth, Industrial Plasma Engineering, Vol 1
- Principles (Institute of Physics Publishing, Bristol, 1995)



\bibitem{Tonks} Lewi Tonks, Phys. Rev. {\bf 97} (6) 1443 (1955)

\bibitem{Corkum} P. B. Corkum, Phys. Rev. Lett. {\bf 31} (13)
809 (1973)

\bibitem{chandrasekhar_60} S. Chandrasekhar, {\it Plasma Physics} (Chicago Press, Chicago, 1960)

\bibitem{Zoran} Z. M. Raspopovi\'{c}, S. Dujko, T. Makabe, and Z.
Lj. Petrovi\'{c}, Plasma Sources Sci. Technol. {\bf 14} 293 (2005)

\bibitem{Zoran1} Zoran Raspopovi\'{c}, Sava Sakad\v{z}i\'{c}, Zoran
Lj. Petrovi\'{c} and Toshiaki Makabe, J. Phys. D: Appl. Phys. {\bf
33} 1298 (2000)

\bibitem{White} R. D. White, R. E. Robson, K. F. Ness and T.
Makabe, J. Phys. D: Appl. Phys. {\bf 38} 997 (2005)

\end{thebibliography}

\end{document}